Intergranular corrosion in evolving media: experiment and modeling by cellular automata

S. Guiso[a], N. Brijou-Mokrani[a], J. de Lamare[a], D. Di Caprio[b], B. Gwinner[a], V. Lorentz[a] and, F. Miserque[a],

[a] Université Paris-Saclay, CEA, Service de la Corrosion et du Comportement des Matériaux dans leur Environnement, 91191, Gif-sur-Yvette, France
[b] Chimie ParisTech, PSL Research University, CNRS, Institut de Recherche de Chimie Paris (IRCP), F-75005 Paris, France

**Abstract**

We investigate the impact of the oxidizing character of the nitric medium on the evolution of the intergranular corrosion of a stainless steel. In two different oxidizing conditions ("severe" and "soft"), the corroded surface of the steel reaches a different steady state: the oxide is thicker, the intergranular grooves are thinner and the surface area is larger in the "severe" conditions than in the "soft" ones. Then we investigate the effect of an alternative "severe then soft" corrosion sequence. We show that the system re-adapts to the "soft" conditions without memory effect from the previous "severe" ones.

**Keywords**

Austenitic stainless steel, Cellular automata modeling, Intergranular corrosion, Nitric acid, X-ray photoelectron spectroscopy

**Introduction**

Grain boundaries in materials are often considered as a weakness regarding the chemical attack by the environment. The preferential attack of these grain boundaries results in a specific corrosion mode named intergranular corrosion (IGC). Challenges relative to intergranular corrosion concern various materials such as nickel based alloys [1-3], aluminum based alloys [4-6] or steels [7-9]. One specific issue is about IGC of non-sensitized stainless steels (SSs) in nitric acid. This is mainly linked to the domain of the spent nuclear fuel reprocessing, where SSs are used for containing nitric acid solutions at various concentrations and temperatures [10-12]. IGC of non-sensitized SSs is induced by a differential of reactivity between grains and grain boundaries. This is due to the fact that grain boundaries are crystallographic defects [3, 13, 14] and are prone to segregation of minor elements such as phosphorous, sulfur or silicon [13-23] that both increase the grain boundaries dissolution compared to the grains themselves. Therefore, grain boundaries are preferentially dissolved compared to the grains themselves. Consequently, grooves are formed along the grain boundaries, which progression in the SS generates a periodic detachment of grains.

Over the last fifteen years, IGC of non-sensitized SSs has been investigated through both experimental and modeling approaches. From the experimental point of view, different techniques are used to characterize IGC. In addition to qualitative observations of the surface, optical or scanning electron microscopy can be performed on the cross section of the sample to geometrically characterize the intergranular grooves in 2D [15, 24-33]. A promising 3D characterization approach has also been attempted by µ-Xray-tomography [34]. The measurement of the mass loss as the function of time gives also important information regarding the kinetics of IGC [18, 24, 26, 29-31, 35-41]. Sometimes, the chemical characteristics of the surface are analyzed by X-ray photoelectron spectroscopy (XPS) [26, 38, 42, 43]. An attempt to characterize IGC with ultrasonic methods was proposed by Jothilakshmi *et al.* [32]. In addition to this experimental information, different models were recently proposed to simulate IGC of non-sensitized SSs in nitric acid. A simple analytical approach was proposed for simulating the evolution of the mass loss [44]. A geometrical approach in 2D was developed for simulating the evolution of the material/solution interface morphology during IGC [31, 44]. More realistic models were also developed using the method of cellular automata (CA) for simulating the geometrical evolution of the material in a 3D realistic granular structure [15, 41, 45-48].

All these experimental and modeling approaches allowed studying the influence of different metallurgical and chemical parameters on IGC. It appears that IGC could be dependent on the surface exposed to corrosive medium regarding the rolling direction [29, 30, 39, 49], on the chemical composition of the SSs, which in turn influence the chemical composition of the grain boundaries [15, 18, 24, 25, 30, 38, 42, 43, 50, 51], on the grain size [40] and on the metallurgical treatment [35]. IGC of SSs was also studied in a pure nitric medium at different concentrations [38, 42] but also with the addition of different oxidizing species at different levels [36, 37, 49, 51]. However, these parameters were only correlated to the corrosion potential and corrosion kinetics, without further information about the material/solution interface in terms of chemical composition and morphology. In this paper, we intend to fill this gap by investigating IGC of a SS in two different nitric acid media. We investigated the influence of the chemical medium nature on the elemental composition of the oxide layer (analyzed by XPS) and the geometrical characteristics of the solid/solution interface. For this last one, as it was difficult to extract information from experiments, we used CA modeling. Indeed, we showed that CA is able to reproduce accurately the experimental behavior [33, 47].

Moreover, we also investigated cases where the corrosive environment evolves over time during a given experiment. The objective was to study how the IGC of a given SS sample may be influenced by the fact that this SS was previously corroded in a first medium. To our knowledge, this aspect has never been studied despite it may concern many industrial issues, for example: the case of devices whose corrosive environment evolves in time or the case of the alternation of phases of rinsing (with one first chemical medium) and working (in a second chemical medium).

## Experimental and modeling approaches

### Experimental conditions and characterization methods

We performed corrosion tests (electrochemical measurements and immersion tests) with an AISI 310L SS provided by Creusot Loire Industrie. This stainless steel has a low carbon content and was solution annealed and quenched. The chemical composition is given in Table 1. We used the standard NF EN ISO 643:2003 to determine the mean grain size (84 µm).

Table 1. Chemical Composition of AISI 310L SS (in wt.%).

| Fe | C | Cr | Ni | Mn | Si | P | S | Mo | Nb |
|---|---|---|---|---|---|---|---|---|---|
| bal. | 0.006 | 24.32 | 21.13 | 1.03 | 0.13 | 0.016 | 0.001 | 0.08 | 0.115 |

For the corrosion tests, we used nitric acid at 8 mol.l$^{-1}$ at boiling point temperature (111 °C). In some cases, we added 150 or 300 mg/L of vanadium(V) (under the form of $V_2O_5$). By introducing or not vanadium(V), we intended to impose different oxidizing conditions to the material regarding the difference of standard potential in nitric acid (E°($NO_3^-$/$HNO_2$) = 0.934 V/SHE at 25 °C [52]) and in presence of vanadium(V), (E°($VO_2^+$/$VO^{2+}$) = 1.004 V/SHE at 25 °C [52]). For the rest of the paper, we name "soft" conditions the experiments without vanadium(V) and "severe" conditions the ones with additions of vanadium(V).

We performed electrochemical measurements (open circuit potential and linear voltammetry measurements) using a classical 3-electrodes configuration. The working and the counter electrodes were a 310L bullet shaped electrode (surface area 2 mm$^2$) and a platinum basket, respectively. The reference electrode was a mercury/mercurous sulfate electrode (Hg/Hg$_2$SO$_4$ - MSE, E = +0.65 V/SHE at 25 °C). The electrochemical measurements were performed using a VSP workstation controlled by the software EC-Lab v.10.37 (Biologic).

The conditions and durations of the 3 immersion tests are given in Table 2. The SS samples had a parallelepiped shape of dimension 30 × 20 × 1.5 mm$^3$. We degreased them with an ethanol/acetone mixture in an ultrasonic bath. We pickled them in a mixture (600 mL of 37 %wt. HCl + 300 mL of 52,5 %wt. HNO$_3$ + 10 mL of 40 %wt. HF + 90 mL of H$_2$O) during 15 minutes so as to remove a 20 µm metal thickness. Three specimens were immersed in a 1 L glass reactor so that the metallic surface area on the solution volume S/V ratio was 0.4 dm$^2$/L. We removed the specimens at the end of each period. The samples were then rinsed with demineralized water and ethanol, dried with compressed air and weighted (AT 20 balance model from Mettler-Toledo, precision 0.01 mg). The solution was renewed after each corrosion period. To characterize the mean corroded thickness (that is on average over the whole corroded surface), we expressed the sample mass loss as a function of time in terms of an equivalent thickness loss:

$$\Delta e = \Delta m / ( \rho \times S ) \times 10^4 \qquad \text{Equation 1}$$

With  $\Delta e$ the equivalent thickness loss (in µm)

$\Delta m$ the measured mean mass loss (in g)

$\rho$ the SS density (8 g/cm$^3$ approx.).

$S$ the sample surface (13.5 cm$^2$)

This equivalent thickness loss does not take into account the heterogeneous character of IGC. To investigate the evolution of the material/solution interface (global morphology, grooves at grain boundaries) we cut one of the samples at regular time intervals and observed the cross-sections with a microscope (model GX51 from Olympus).

Table 2: Conditions of the immersion corrosion tests. In tests n°1 and n°2, the corrosive medium remained constant during the test. In test n°3, the corrosive medium was modified after 1935 h.

| Test | | Conditions | Duration |
|---|---|---|---|
| Test n°1 | | "Severe" | $t_{\text{severe}} = 2564$ h |
| Test n°2 | | "Soft" | $t_{\text{soft}} = 16815$ h |
| Test n°3 | Phase n°1 | "Severe" | $t_{\text{severe}} = 1935$ h |
| | Phase n°2 | "Soft" | $t_{\text{soft}} = 16815$ h $(t_{\text{test3}} = t_{\text{severe+soft}} = 17783$ h$)$ |

For surface chemical analysis, we performed XPS measurements. The device and the conditions of use have been already described in reference [53]. For the determination of relative proportions of oxide and metallic contribution, Ni-2p$_{3/2}$, Cr-2p$_{3/2}$ and Fe-2p$_{3/2}$ spectra have been deconvoluted. The main contributions used are presented in Table 3. Several arbitrary contributions are necessary to fit the iron and chromium oxide due to a multiplet splitting making the spectral analysis particularly complex [54, 55]. Nickel element is observed only in the metallic chemical state.

As XPS is an ex-situ measurement, the analysis of the oxide by XPS could be influenced by the previous steps of rinsing with water, drying with compressed air and storage, after immersion in nitric acid. In spite of this possible bias, it has been previously shown that the XPS analysis is representative of the state (passive/transpassive) in which the SS is polarized in nitric acid [26].

Table 3: Parameters used for the deconvolution of Cr, Ni and Fe-2p$_{3/2}$ core level spectra (Avantage$^{TM}$ software) binding energies and full width at half maximum (FWHM).

|  | Ni-2p$_{3/2}$ | Fe-2p$_{3/2}$ | | | Cr-2p$_{3/2}$ | | | |
|---|---|---|---|---|---|---|---|---|
| **Chemical state** | Metallic | Metallic | Oxide | | Metallic | Oxide | | |
|  |  |  | Peak 1 | Peak 2 |  | Peak 1 | Peak 2 | Peak 3 |
| **Binding Energy (eV)** | 853.0 ± 0.3 | 853.0 ± 0.3 | 709.8 ± 0.3 | 712.0 ± 0.3 | 853.0 ± 0.3 | 576.1 ± 0.3 | 577.3 ± 0.3 | 578.5 ± 0.3 |
| **FWHM (eV)** | 1.4 | 1.0 | 3.0 | 3.3 | 1.5 | 1.7 | 1.8 | 1.9 |

Cellular Automata model

The CA automata model used in this paper is identical to the one developed in a previous work. Only the main elements are recalled here. All details are given in the reference [33].

The CA approach consists first in discretizing a volume (box) into a 3D hexagonal grid of cells. The grid taken for the simulation is a box of 1280x1280x1280 cells. In the grid, a cell can belong to a grain (GRN state), a grain boundary (IGN state) or the solution (SOL state). The material granular microstructure (GRN and IGN cells) is simulated numerically by a Voronoi diagram (Figure 1). The procedure consists in fixing a number of seeds N$_{seeds}$ randomly distributed in the grid. Then, virtual spheres centered on these seeds grow progressively. The grain boundaries are generated at the intersection of these spheres to obtain the Voronoi diagram. Consequently, the mean grain size in the model $D_{sim}$ is determined by the number of seeds $N_{seeds}$ chosen for the generation of the Voronoi diagram. The grain size $D_{sim}$ (in number of cells) is estimated from the equation:

$$D_{sim} = \sqrt[3]{\frac{N_{XYZ}}{N_{seeds}}}$$

where $N_{XYZ}$ is the total number of cells in the system. For the simulations, $N_{XYZ}$ and $N_{seeds}$ were fixed to $1280 \times 1280 \times 1280$ and 104, respectively. Consequently, the mean grain size $D_{sim}$ is 272 cells. Considering the scale transformation of 0.309 µm/cell, the mean grain size $D_{sim}$ (in µm) is 84 µm, in accordance with the experimental value. Additionaly, the simulation box of 1280x1280x1280 cells corresponds to a surface dimension of 13.5 mm² and a depth of 395 μm.

Once the Voronoi diagram is generated, the SOL state is assigned to the three uppermost *xy*-layers (Figure 1). Then, the step of simulation of IGC can begin. The corrosion progresses from the top of the grid to the bottom. The differential reactivity of grains and grain boundaries is taken into account in the CA model by introducing to different corrosion probabilities for grains ($P_{grn}$) and grain boundaries ($P_{ign}$), respectively. To simulate adequately IGC, $P_{ign}$ is chosen higher that $P_{grn}$. These probabilities drive the evolution of the system from a time $t$ to the following time $t + \Delta t$, according to transition rules: a IGN (or GRN) cell that is contact with a SOL cell has a probability $P_{grn}$ (or $P_{ign}$) to convert into a SOL cell. The way to fix the values of $P_{ign}$ and $P_{grn}$ so as to reproduce adequately the experimental results (with CA to experiments conversion factors) is described in reference [33].

For the code, we used the C language in the CUDA environment. We performed numerical simulations with NVIDIA Tesla K80 in Dell PowerEdge C4130 servers using Intel Xeon E5-2640 processors. In the case of a (XYZ dimensions) 3D grid, it takes from one to a few days for a simulation run for a single experiment. It depends on the values considered for $P_{grn}$ and $P_{ign}$: the more important they are, the shorter the simulation is. The results are averaged over hundred different Voronoï structures and the error bars given further in the paper represent their dispersion.

From the sample simulated in 3D and corroded at a time *t*, we can estimate the mass loss (as described in reference [33]) and extract virtual 2D cross-sections as illustrated in Figure 5 and Figure 7. On these cross-sections, the mean groove angle $\alpha'$ (see reference [33] for the method), as well as the relative surface in contact with the corrosive medium (see appendix 1 and 2) can be determined. Note that the notation $\alpha'$ refers to the angle determined in 2D from virtual cross-sections of the 3D grid and that its distribution is slightly different from the 3D angles $\alpha$ distribution as discussed in [33, 34].

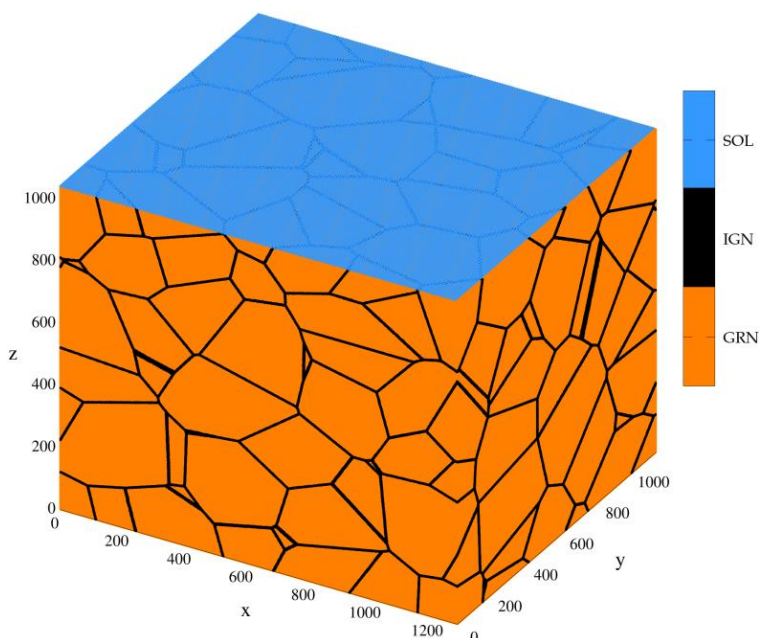

**Figure 1: Example of a 3D initial grid considered for the IGC model obtained numerically by a Voronoï diagram (GRN state cell in orange, IGN state cell in black and SOL state cell in blue). Coordinates are given in number of cells (scale transformation: 0.309 µm/cell).**

## Results

Electrochemical measurements in "severe" and "soft" conditions

We performed preliminary electrochemical measurements in order to specify the conditions of corrosion in both configurations "severe" and "soft". Figure 2 presents the anodic linear polarization of the 310L SS in nitric acid $HNO_3$ 8 mol.l$^{-1}$ at boiling point temperature. Based on the results obtained for a similar system [26], we qualitatively extrapolated the anodic curve below the corrosion potential $E_{corr}$ (that is not directly measurable by electrochemistry because of the cathodic reactions) in order to visualize the respective position of the passive and transpassive domain of the SS. From this linear polarization in "soft" conditions, we observe that the SS is polarized ($E_{corr}$ ~1.11 V *vs.* SHE) in the transition of the passive to transpassive domains, where IGC is expected. Using the Tafel method, the linear part of the anodic curve between 1.12 to 1.15 V vs. NHE can be extrapolated to $E_{corr}$ to estimate an order of magnitude of 1.8 µA.cm$^{-2}$ for the corrosion current density in "soft" conditions. The Faraday's law allows converting this value of corrosion current density into a value of dissolution rate - about 14 µm.y$^{-1}$.

With the addition of vanadium(V), $E_{corr}$ is shifted to a higher anodic potential ($E_{corr}$ ~ 1.14 V *vs.* SHE) in the transpassive domain, where IGC is also expected. As it has been shown that the anodic curve of SS is not significantly influenced by the presence of oxidizing ions

(exempted for Cr(VI)) [10, 11, 53], the corrosion current density in "severe" conditions can be estimated by taking the value of the current density of the anodic curve in $HNO_3$ 8 mol.l$^{-1}$ at this potential of 1.4 V vs. NHE (Figure 2). A value of about 10 µA.cm$^{-2}$ is obtained, which corresponds to a dissolution rate of about 82 µm.y$^{-1}$ (Faraday's law).

From these electrochemical measurements, we showed that the SS is polarized in both conditions in the transpassive domain, where IGC is expected. Moreover, we confirmed the expected effect of vanadium(V), which makes the corrosion conditions more "severe", with a higher corrosion potential and a higher corrosion current density. The objective of the following parts is to characterize IGC in both conditions.

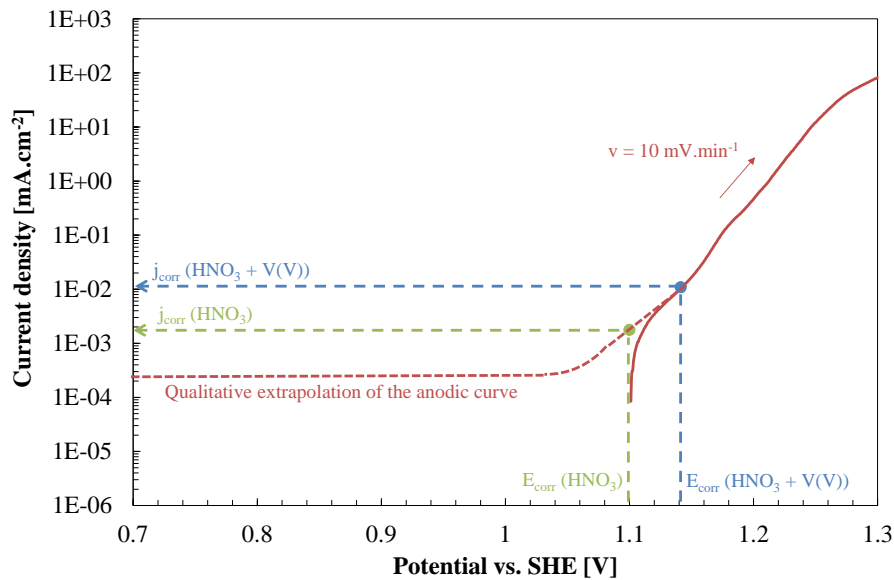

**Figure 2: Anodic linear polarization curve of 310L SS in $HNO_3$ 8 mol.l$^{-1}$ at boiling point temperature (red continuous line). The red dashed line represents the qualitative extrapolation of the anodic curve in the passive domain. The characteristics ($E_{corr}$ and $j_{corr}$) of the system are in green. Those of the system Uranus 65 in $HNO_3$ 8 mol.l$^{-1}$ + V(V) 150 mg.l$^{-1}$ are in blue.**

Immersion tests in stationnary conditions ("severe" and "soft" conditions)

In this part, we discuss the case IGC in both oxidative conditions, respectively, when the chemical medium remains the same as a function of time. Figure 3 presents the observation of the samples before (a) and after the corrosion tests in "severe" (b) and "soft" (c) conditions. In both conditions, samples are darker after corrosion then indicating an effect of the corrosion on the surface. Moreover, the aspect of samples is different between the "severe" (dark grey) and the "soft" (light grey) conditions. This shows that the aspect of the corroded SS is also dependent on the conditions of corrosion. The origin of the difference of aspect is further discussed in terms of chemical composition and IGC morphology of the solid interface with solution.

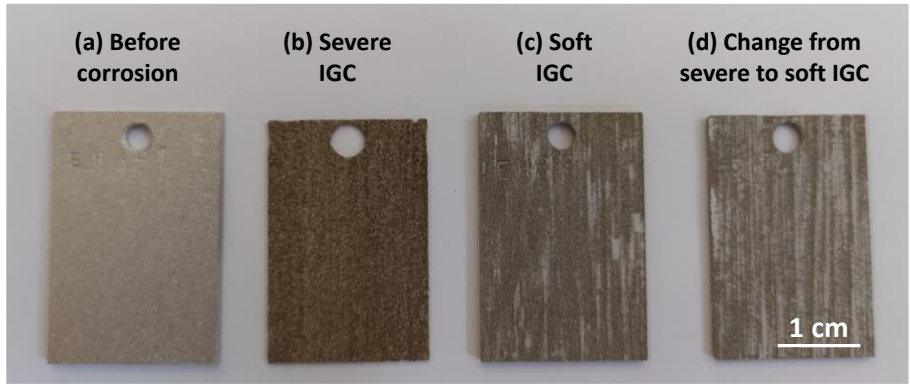

**Figure 3: Aspect of the samples before (a) and after the corrosion tests in the "severe" ($t_{severe}$ = 2564 h) (b), "soft" ($t_{soft}$ = 16815 h) (c) and change from "severe" ($t_{severe}$ = 1935 h) to "soft" ($t_{soft}$ = 16815 h, *i.e.* $t_{severe+soft}$ = 17783 h) conditions (d).**

SEM images of the corroded samples after the immersion tests are given in Figure 4. The aspect of the surface is quite similar for the different conditions. In all cases, a preferential attack is observed along grain boundaries. This intergranular attack has generated a dropping of grains. A localized attack is also observed inside the grains themselves as illustrated in Figure 4(a). It is probably due to the presence of defects in the grains. Indeed, linear (dislocations) or planar (twin boundaries…) defects can also be preferentially attacked like grain boundaries. This localized intragranular attacked seems to be more pronounced in "severe" conditions.

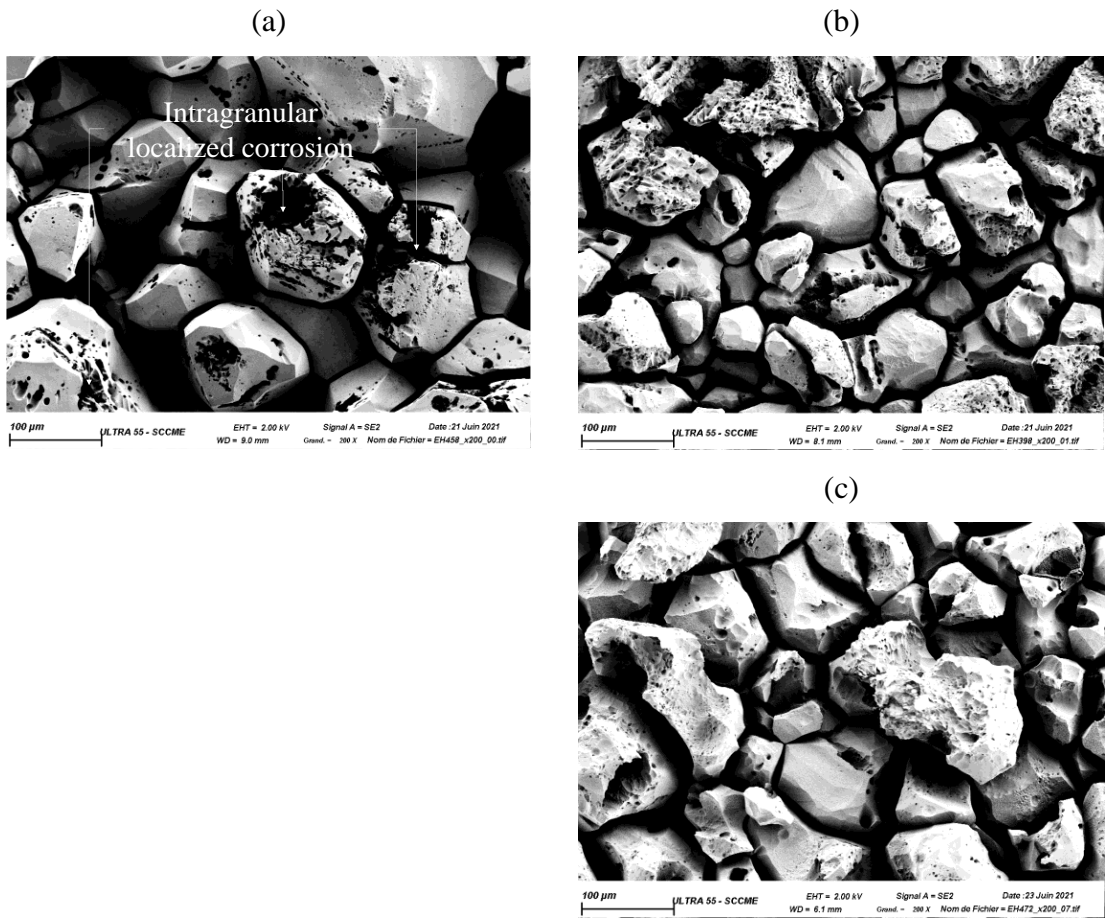

**Figure 4: Observation of the samples by Scanning Electron Microscope after the corrosion tests in the "severe" ($t_{severe}$ = 2564 h) (a), "soft" ($t_{soft}$ = 16815 h) (b) and change from "severe" ($t_{severe}$ = 1935 h) to "soft" ($t_{soft}$ = 16815 h, i.e. $t_{severe+soft}$ = 17783 h) conditions (c).**

The corrosion kinetics is given in terms of morphology (Figure 5 (a) for "severe" and Figure 7 (a) for "soft" conditions) and mass loss (expressed in terms of equivalent thickness loss as described in the experimental section) and corrosion rate (Figure 6 for "severe" and Figure 8 for "soft" conditions). The behavior is typical of the IGC observed for non-sensitized stainless steels and has been extensively discussed in references [26, 31, 44]. IGC begins with the creation of tight grooves at grain boundaries (80 h and 240 h in Figure 5 (a)) that progressively penetrate the steel. This leads to an increase of the SS/solution interface area, which is correlated to the corrosion rate increase as estimated by mass loss (Figure 6). This also leads to the detachment of grains that progressively affects the whole surface (2484 h in Figure 5 (a)) increasing the corrosion rate even more. When the grains detachment has affected the whole surface, the system reaches a corrosion steady state. The surface area does not evolve anymore, since the progression of grooves (increasing the SS/solution interface area) is counter-balanced by the grain detachment (decreasing the SS/solution interface area).

The phenomenology is similar for both oxidizing conditions, but the kinetics is different. It has been shown that this kind of IGC can be described by two different corrosion rates: $V_{grn}$ and $V_{ign}$ which correspond to corrosion rates of the grains and grain boundaries respectively. From the kinetics of IGC (Figure 6 and Figure 8), it is possible to estimate the values of $V_{grn}$ and $V_{ign}*$ (corresponding to the projection of $V_{ign}$ on the vertical axis) by a semi-empirical approach presented in [44]. $V_{ign}$ is deduced from $V_{ign}*$ using a tortuosity factor $k$: $V_{ign} = V_{ign}* \times k$, with $k = 1.18$ [33]. Results are given in Table 4. Both corrosion rates are faster in "severe" than in "soft" conditions. Note that the values of the dissolution rate $V_{grn}$ are in relative agreement with those estimated from the corrosion current density measurement (Figure 2). This corroborates that $V_{grn}$ controls the dissolution of the SS (which can be measured by electrochemical measurement), whereas $V_{ign}$ controls the dropping of grains (which has no influence on the anodic current measured) [44].

The transient time necessary to reach the steady state can be estimated as follows [44]:

$$\frac{3}{2}\frac{D}{V_{ign}} \qquad \text{Equation 2}$$

where $D$ represents the average grain diameter (84 µm). Therefore, the transient time is lower in "severe" conditions (Table 4). Durations of the experiments were fixed so that the steady state could be almost reached in both cases, that is 2564 h for "severe" and 16815 h for "soft" conditions, respectively.

Table 4. Corrosion rates and transient time for the "severe" and "soft" IGC cases

| IGC | $V_{grn}$ / µm.y$^{-1}$ | $V_{ign}*$ / µm.y$^{-1}$ | $V_{ign}$ / µm.y$^{-1}$ | Transient time / h |
|---|---|---|---|---|
| "severe" | 273 | 1656 | 1954 | 565 |
| "soft" | 12 | 56 | 66 | 16703 |

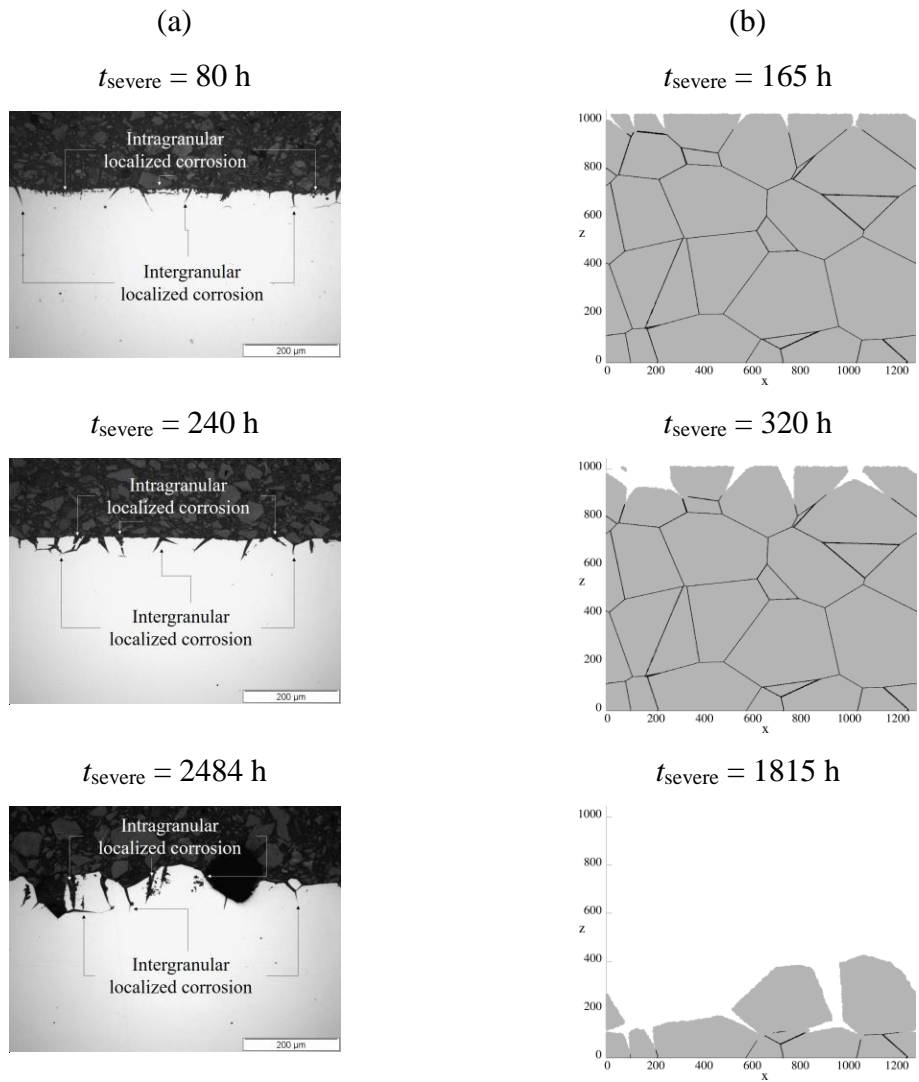

**Figure 5**: Evolution of the IGC observed on cross-sections in the case of a "severe" IGC. (a) Experiment – scale given in µm (b) CA simulations at different times - coordinates given in number of cells (scale transformation: 0.309 µm/cell).

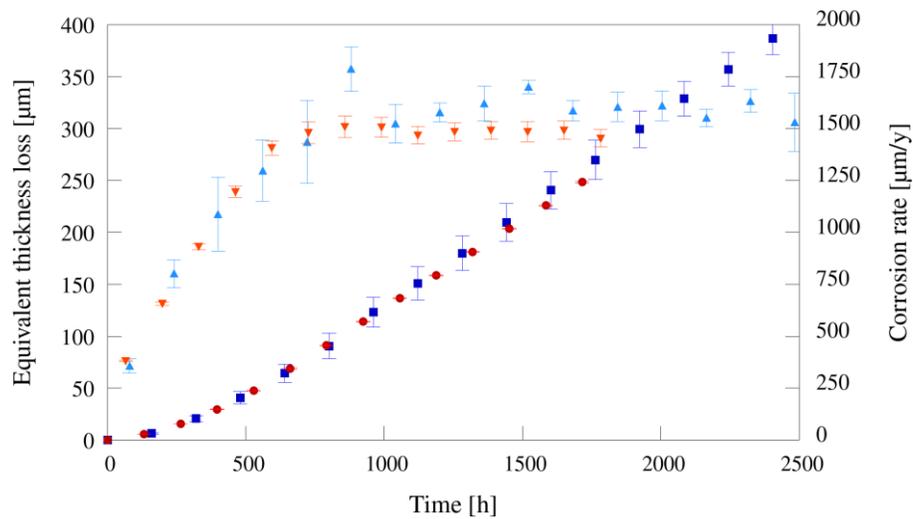

**Figure 6: Comparison between experimental results (in blue, error bars indicate the dispersion of results on 3 samples) and CA simulations (in red, error bars indicate the dispersion of simulations over hundred different Voronoï structures) in the case of a "severe" IGC. Squares represent the equivalent thickness loss and triangles the corrosion rate.**

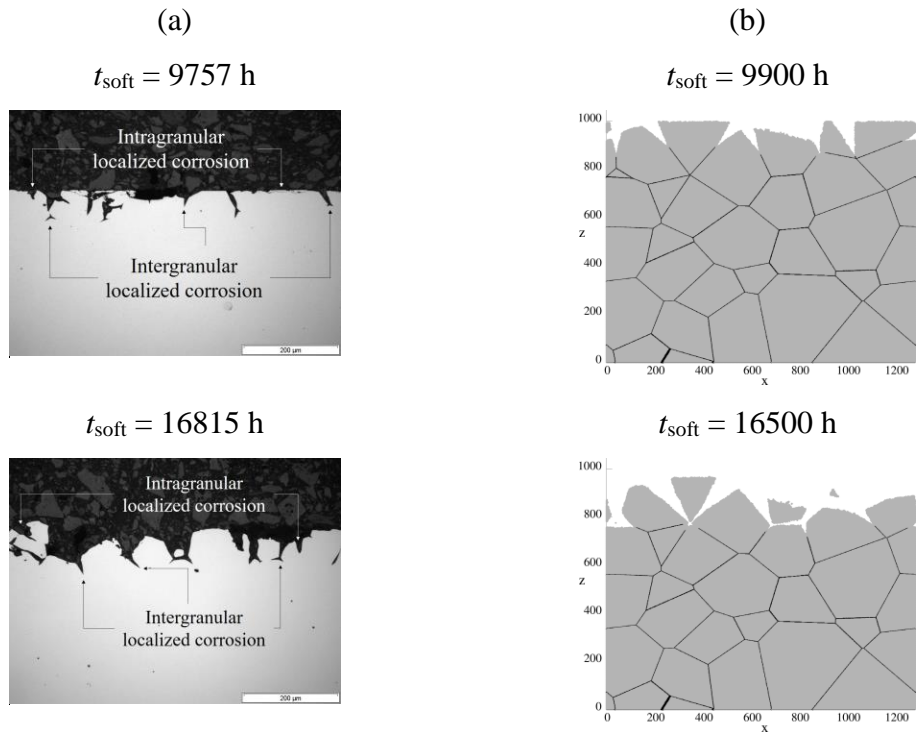

**Figure 7: Evolution of the IGC observed on cross-sections in the case of a "soft" IGC. (a) Experiment (b) CA simulations at different times.**

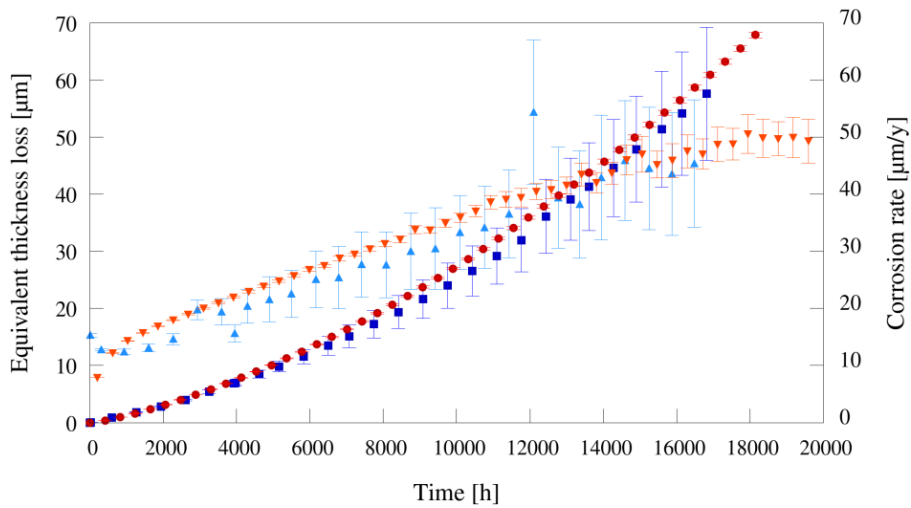

**Figure 8: Comparison between experimental results (in blue, error bars indicate the dispersion of results on 3 samples) and CA simulations (in red, error bars indicate the dispersion of simulations over hundred different Voronoï structures) in the case of a "soft" IGC. Squares represent the equivalent thickness loss and triangles the corrosion rate.**

In order to get a quantitative description of the morphology, we used a CA modeling of the experiments. In a previous work [33], we showed that this model reproduces accurately the morphological evolution of IGC. This CA numerical approach enables to quantify precisely the morphological characteristics of the system, in terms of grooves angles $\alpha'$ distribution and the surface in contact with the solution. These parameters would be difficult to quantify or be only partially accessible by an experimental approach, which justifies the use of this CA modeling. The time scaling factor is chosen in order to set and maximize the parameter $P_{ign,severe} = 1$ and thus minimize the duration of the calculations. The input corrosion probabilities are given in Table 5 for both IGC conditions. Moreover, conversion factors were also used in order to convert space and time scales from CA modeling to real experiments [33]: the space scaling factor A and time the scaling factor B were fixed to 0.31 µm/cell and 0.825 h/iteration, respectively.

CA simulated time evolutions of the mass loss (as expressed in terms of equivalent thickness loss) are given in Figure 6 and Figure 8. As exhaustively discussed in [33], CA simulations are able to reproduce accurately the IGC kinetics observed experimentally. They show the initial transient regime characterized by a gradual increase in the corrosion rate. The following constant rate corresponds also accurately to the experimental one. 2D cross sections were extracted from the 3D Voronoi structures at different corrosion times (Figure 5 and Figure 7). The evolution of the surface morphology is well reproduced by the CA simulations, with the presence of intergranular grooves. These grooves progress in the SS and lead to grain dropping. The main experiment/model difference is about localized intragranular corrosion, that is clearly evidenced experimentally (Figure 4, Figure 5(a) and Figure 7(a)). This localized intragranular corrosion has an irregular form and is due to the attack of grain structural defects, such as dislocation (1D defect) or twin joints (2D defect). At present time, the CA model does not take into account this kind of localized intragranular attack. Thus, the following quantitative analysis concerns only the intergranular grooves. Note that the mass loss predictions of the CA model are accurate, since the mass loss at steady state is driven by grain dropping (that is governed by $P_{ign}$).

As discussed above, the quantitative analysis of experimental grooves is difficult and only partial. Therefore, only the grooves from the CA simulations are further discussed. The distribution of the grooves angle $\alpha'$ (in degrees) at steady state of IGC is given in Figure 9. It appears that the grooves are thinner in "severe" (mean value 20.5, standard deviation 5.0) than in "soft" (mean value 26.59, standard deviation 5.9) conditions. Thus, it appears that the more oxidant the corrosive solution, the thinner the grooves and the larger the ratio $V_{ign}/V_{grn}$.

Note that from Beaunier [56, 57], the groove angle $\alpha$ is theoretically related to the $V_{ign}/V_{grn}$ ratio by the relation:

$$\frac{V_{ign}}{V_{grn}} = \frac{1}{sin(\alpha/2)} \qquad \text{Equation 3}$$

This theoretical relation is verified for the CA simulations in both "severe" and "soft" conditions. This thus reinforces the fact that the CA model is able to reproduce IGC in a realistic way.

Table 5: Corrosion probabilities used as input data in the CA model and calculated from the experimental results based on the procedure given in reference [33].

|  | $P_{ign}$ | $P_{grn}$ |
|---|---|---|
| "Severe" | $P_{ign,severe} = 1$ | $P_{grn,severe} = P_{ign,severe} \times \dfrac{V_{grn,severe}}{V_{ign,severe}}$<br>$P_{ign,soft} = 0.140$ |
| "Soft" | $P_{ign,soft} = P_{ign,severe} \times \dfrac{V_{ign,soft}}{V_{ign,severe}}$<br>$P_{ign,soft} = 0.0338$ | $P_{grn,soft} = P_{ign,soft} \times \dfrac{V_{grn,soft}}{V_{ign,soft}}$<br>$P_{grn,soft} = 0.00614$ |

Table 6: Intergranular grooves angles $\alpha'$ characteristics at steady state in the different corrosion conditions (CA simulations)

| Exp., time | Mean value (degrees) | Standard deviation (degrees) | Error | $1/(\sin(\alpha'/2))$ |
|---|---|---|---|---|
| "Severe" condition $t_{severe} = 825$ h | 20.54 | 5.00 | 0.13 | 5.61 |
| "Soft" condition $t_{soft} = 20\,625$ h | 26.59 | 5.89 | 0.16 | 4.35 |
| Change "severe" $t_{severe} = 825$ h | 20.67 | 4.88 | 0.14 | 5.57 |
| to "soft" condition $t_{severe+soft} = 20\,625$ h | 26.04 | 5.92 | 0.17 | 4.44 |

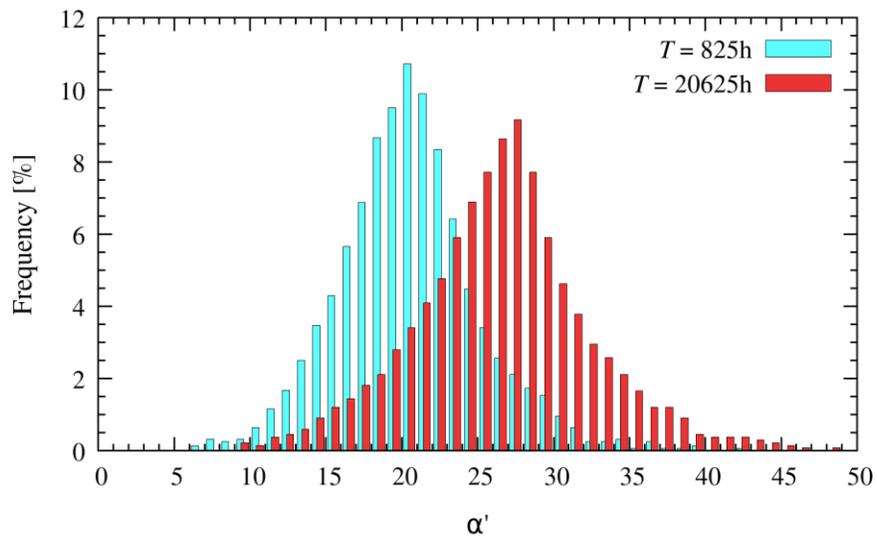

**Figure 9: Distribution of the grooves angles α' in the steady state in "severe" (in blue, evaluated at $t_{severe}$ = 825 h) and in "soft" (in red, evaluated at $t_{soft}$ = 20625 h) conditions**

**(CA simulations). In both cases, histograms are realized with the same classes but are slightly shifted for a better view.**

Moreover, we investigated how the grooves angles distributions affect the surface in contact with the corrosive solution. For this purpose, the relative surface was determined from the CA simulations as described in the appendix 1 and 2. Results are given in Figure 10. In both "severe" and "soft" conditions, the relative surface increases from its initial "flat surface value" (100%) to reach a larger steady state value. This evolution results from the appearance and progression of grooves in the SS that leads to an increase in the surface in contact with the solution, which is progressively counterbalanced by the dropping of grains that gradually affects the whole surface and tends to decrease the contact surface. The constant relative surface steady state regime corresponds to an equilibrium between these two mechanisms. The steady state relative surface value is higher (238.4 ± 0.4%, estimated in the interval $t_{severe}$ = 1568 - 1732 h) for the "severe" conditions than in "soft" conditions (223.7 ± 0.5%, estimated in the interval $t_{soft}$ = 17325 - 20625 h). Therefore, the thinner groove angles, the larger the relative surface. This is in agreement with the results of Gwinner *et al.* obtained with a model based on a 2D geometrical approach [44].

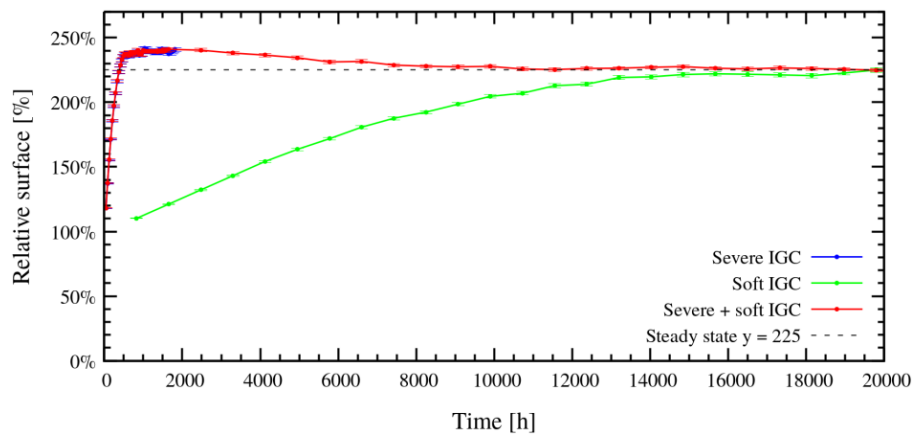

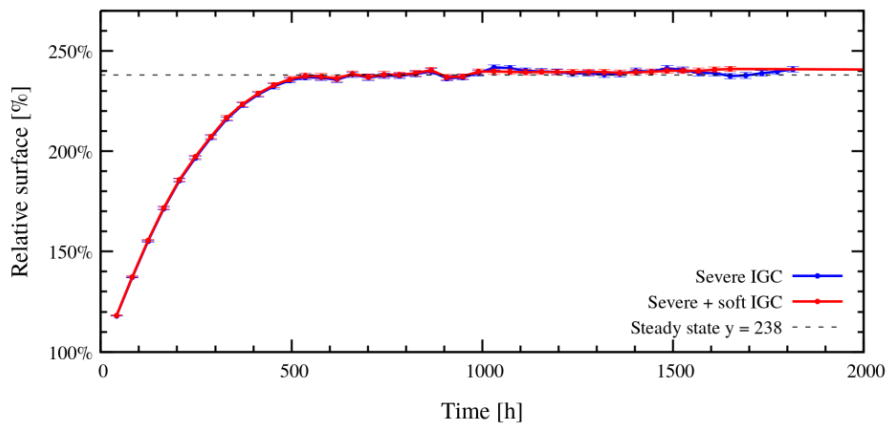

**Figure 10: Relative surface for sections of the lattice for the three IGC corrosion regimes (CA simulations). The bottom plot is a zoom of the top one at short times.**

After having described the system in a morphological point of view, we also analyzed the chemical composition of the surface using XPS relative quantification in both conditions at the end of the experiments (Figure 11). As XPS measurements were performed on very rough surfaces (Figure 10), a quantitative analysis of the results would be tricky as discussed in [58, 59]. Nevertheless, the diameter of the focused X-rays beam is relatively large (900 µm) regarding the order magnitude of roughness (about 100 µm). This should contribute to smooth the effect of roughness. Moreover, the morphology of the surface is quite similar in all conditions (Figure 10) and the XPS measurements were performed in identical conditions (in particular, with the same angle between the surface and the photoelectrons detector/analyzer). Then we can assume that the differences between surfaces are sufficiently small, to allow a relative comparison between XPS results [60].
The XPS response is qualitatively similar in "severe" and "soft" conditions. The spectra show the presence of oxide and metal (associated to the presence of the metal under the oxide layer). The relative proportions of the oxide and metal contributions (for the major elements contained in the 310L SS: Iron, nickel, chromium) are reported in Figure 11 (a). The relative contribution of the oxide is larger in the "severe" than in the "soft" IGC conditions. This indicates that the oxide layer is thicker in the "severe" IGC conditions (with a higher corrosion potential $E_{corr}$ as illustrated in Figure 2) than in the "soft" IGC conditions. This is in accordance with the results of Tcharkhtchi *et al.* who showed an increase of the oxide thickness of a 304L SS in nitric acid with the increase of the corrosion potential $E_{corr}$ [26].

Figure 11 (b) gives the chemical composition of the oxide layer. As classically observed for these SSs corroded in nitric acid media [26, 43, 61], the oxide is mainly composed of chromium with a small proportion of iron. Nickel oxide is not observed. The "severe"/"soft" IGC conditions have a small influence on the chromium/iron ratio: a little larger relative enrichment in chromium is observed for the "severe" condition.

The chemical composition of the underlying metal is given in Figure 11 (c). Its relative composition differs from the bulk material: relatively enriched in nickel, compared to chromium and iron. This observation has already been reported for SSs in nitric acid [26]. It could be due to the fact that nickel is less oxidized than chromium and iron (as illustrated in Figure 11 (b)) and consequently accumulates under the oxide. This nickel enrichment seems to be larger in the "severe" conditions than in the "soft" conditions. This could be linked to the fact that the metallic zone that is analyzed by XPS is thicker (because the oxide is thinner) in "soft" conditions than in "severe" conditions. Therefore, the nickel relative enrichment in the underlying metal is more averaged with the bulk metal composition in the "soft" than in the "severe" conditions.

(a) Relative quantification of the total oxide and the underlying metal contributions (for the major elements contained in the 310L SS: Fe, Ni, Cr)

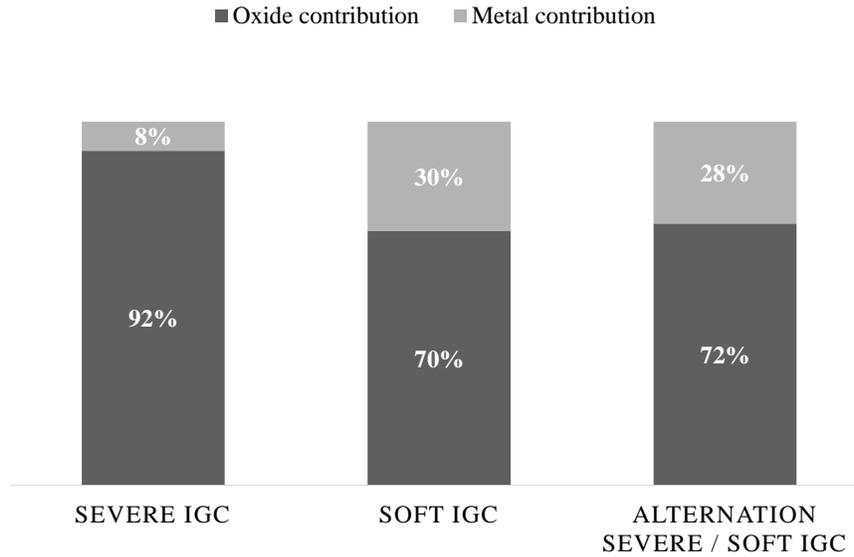

(b) Relative quantification of the composition of the oxide layer

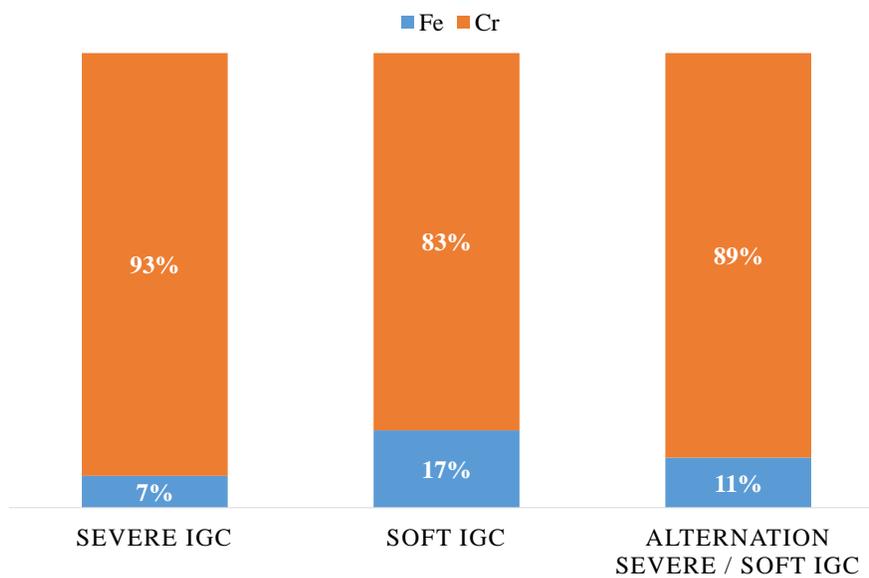

(c) Relative quantification of the composition of the underlying metal

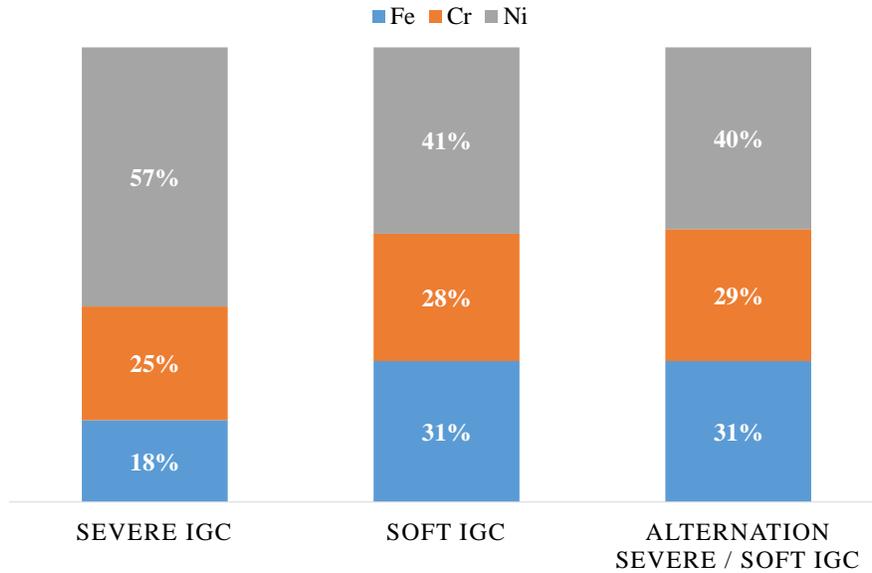

**Figure 11: XPS relative quantification of the chemical characteristics of the surface of 310L SS samples corroded in the "severe" ($t_{severe}$ = 2564 h), "soft" ($t_{soft}$ = 16815 h) and change from "severe" ($t_{severe}$ = 1935 h) to "soft" ($t_{soft}$ = 16815 h, i.e. $t_{severe+soft}$ = 17783 h) conditions.**

We showed that in stationary conditions, the IGC characteristics depend on the oxidizing power of the corrosive medium. In more "severe" conditions, the SS is polarized at higher potential in the transpassive domain which has different consequences on the corrosion of the SS. Firstly, the dissolution of the SS is faster ($V_{grn}$ is larger). Interestingly, the presence of a thicker oxide on the surface is evidenced by XPS. Secondly, the intergranular corrosion is faster ($V_{ign}$ is larger), which leads to a shorter transient time to reach the IGC steady state (in accordance with [35], where the transient time is shown to be inversely proportional to $V_{ign}$). Moreover, the global corrosion rate reached at steady state (which corresponds to $V_{ign}$*) is larger. Thirdly, the ratio $V_{ign}/V_{grn}$ is increased which has the consequence that the grooves angles are thinner. Finally, the total surface reached at steady state is larger.

As discussed in the introduction, it is of practical interest to investigate the influence of changing the nature of the corrosive medium during the life duration. This is the objective of the next part.

Immersion tests with change from "severe" to "soft" IGC conditions

In order to study the effect of a change in the IGC conditions, we firstly corroded samples during 968 h in the conditions of "severe" IGC (with vanadium(V)) until the steady state is reached. Then we changed the solution into "soft" conditions (without vanadium(V)).

Figure 12 and Figure 13 (a) display the experimental evolution of the mass and the morphology as a function of time, respectively. Concerning the first step in "severe" conditions, the experimental behavior is identical to the one observed in "severe" stationary conditions in the previous part, with a progressive increase of the corrosion rate and a similar IGC morphology. At the end of this early stage of 968 h, the steady state is reached in terms of mass loss (Figure 12) and surface morphology (Figure 13 (a)).

When the medium is changed into "soft" conditions, the corrosion behavior changes (Figure 12 and Figure 13 (a)). For an easier interpretation of the results, the evolution of the corrosion rate is given in Figure 14 and compared to the one in "soft" conditions only discussed in the previous part. This allows to observe directly the IGC evolution in the "soft conditions", when the samples are initially flat and not corroded and when the samples are pre-corroded in the "severe conditions". In the case of pre-corroded samples, the corrosion rate suddenly decreases in accordance with the new "soft" IGC conditions (Figure 12 and Figure 14). However, it does not reach immediately a constant value. It slowly decreases as a function of time even if the solution (that means the corrosiveness) remains the same. The system reaches a new steady state, which is identical in terms of corrosion rate (about 50 $\mu m.y^{-1}$) to the case of non pre-corroded samples. Moreover, the steady state is reached at the same time in both cases. This transient time is in accordance with the one estimated by Equation 2. This corresponds to the time necessary to remove a thickness equivalent to the mean grain size. Thus, whatever the history of the SS (already corroded in "severe" IGC conditions or not), the corrosion needs to consume the equivalent of a layer of grains to erase the effects of this history.

Figure 13 shows the evolution of the IGC morphology from the "severe" conditions at steady state (at $t_{severe}$ = 968 h) to the "soft" conditions (until $t_{severe+soft}$ = 17783 h). A slow evolution is observed between a highly rough surface (resulting from the "severe" IGC conditions) to a less rough surface (resulting from the "soft" IGC conditions). In the same way, SEM observations in Figure 4 show that the "soft" conditions tend to smooth the serious intragranular corrosion induced by "severe" conditions. CA simulations were performed to investigate quantitatively this point. The same input corrosion probabilities were considered as for the previous simulations (Table 5). The first step was first modeled in "severe" conditions during 990 h. Then the probabilities were changed into "soft" conditions to model the rest of the simulation.

Simulations reproduce correctly the mass loss kinetics (Figure 12) and the evolution of the morphology (Figure 13) observed experimentally. It can be deduced that in addition to being able the simulate IGC in stationary conditions (previous section), the model is able to reproduce accurately the IGC behavior when the corrosion conditions evolve in time.

Figure 15 (b) shows that the characteristics of the simulated grooves angles evolve from relatively thin angles under "severe" conditions to larger angles in "soft" conditions. Consequently, the relative surface in contact with the solution decreases slowly from a value of 238.7 ± 0.7% (estimated in the interval $t_{severe}$ = 1568 – 1733 h) in "severe" conditions to about 225.4 ± 0.5% (estimated in the interval $t_{severe+soft}$ = 17325 - 20625 h) in "soft" conditions, the same value as the experiment in steady "soft" conditions.

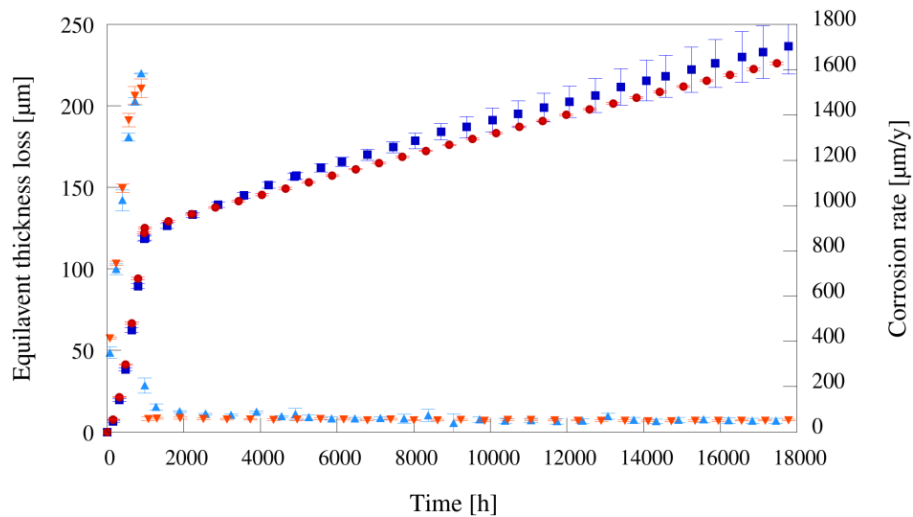

**Figure 12: Comparison between experimental results (in blue, error bars indicate the dispersion of results on 3 samples) and CA simulations (in red, error bars indicate the dispersion of simulations over hundred different Voronoï structures) in the case of alternation a "severe" and a "soft" IGC. Squares represent the equivalent thickness loss and triangles the corrosion rate.**

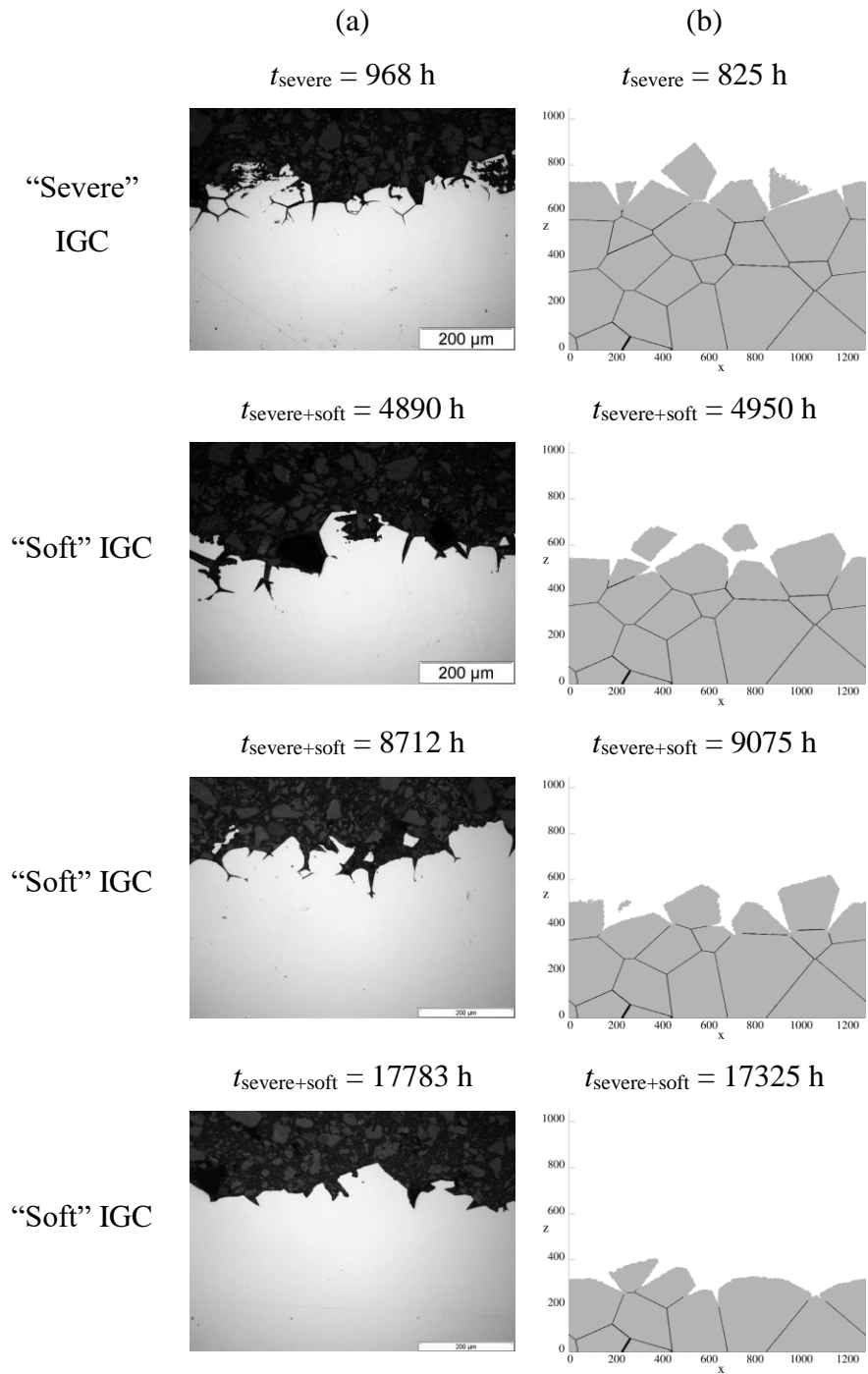

**Figure 13: Evolution of the IGC observed on cross-sections in the case of alternation a "severe" and a "soft" IGC. (a) experiment and (b) CA simulations.**

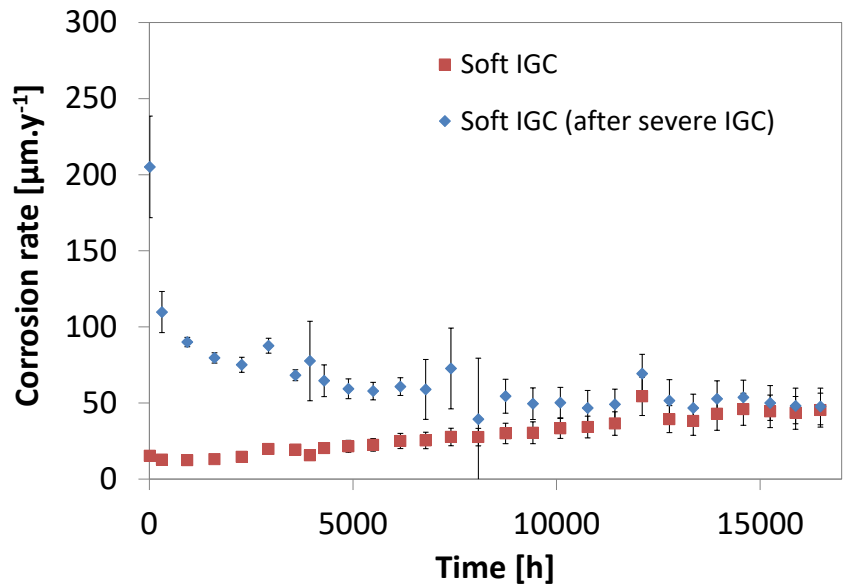

**Figure 14: Comparison of the corrosion rate estimated for "soft" IGC only and "soft" IGC after "severe" IGC (experimental results). Error bars indicate the dispersion of results on 3 samples.**

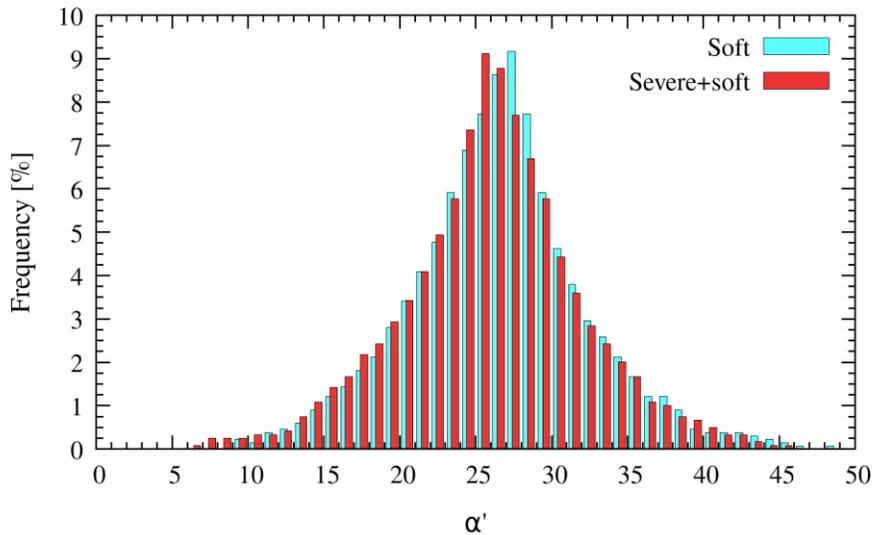

**Figure 15: Distribution of the grooves angles $\alpha'$ in the steady state in "soft" (in blue, evaluated at $t_{soft}$ = 20625 h) and in change from "severe" to "soft" (in red, evaluated at $t_{severe+soft}$ = 20625 h) conditions (CA simulations). In both cases, histograms are realized with the same classes but are slightly shifted for a better view.**

The characteristics of the oxide layer on the surface also re-adapts to the new "soft" conditions (Figure 11). In particular, the apparent oxide thickness is similar in "soft" conditions and in change from "severe" to "soft" conditions (Figure 11 (a)).

Because of the oxide and morphology re-adaptation to the new "soft" conditions, the aspect of the surface also changes from dark grey (characteristic of the "severe" conditions) to light grey (characteristic of the "soft" conditions) as shown in Figure 3.

## Interest and perspectives of the CA model

The work described above has shown that the CA model is able to simulate accurately the IGC behavior in both stationary and non-stationary chemical conditions. Thus it could be used to obtain long-term predictions about IGC in cases where devices are exposed to time-dependent corrosive environments. Suppose, as an example, a time-dependent medium where the concentration of the corrosive specie is bounded by $C_{min}$ and $C_{max}$ (Figure 16(a)). In order to model the IGC behavior of such a system, the following procedure is proposed (illustrated in Figure 16(b) and Figure 16(c)). Two short time immersion tests can be performed in stationary conditions at $C_{min}$ and $C_{max}$, respectively. At the end of the experiment ($t_{exp}$), the morphology of the IGC grooves can be characterized in terms of groove angles and deepnesses from the observation of the cross section of the corroded samples. As described in reference [44], both corrosion rates $V_{grn}$ and $V_{ign}$ can be estimated from both these geometrical parameters. Then, following the same approach as presented in this paper (estimation of corrosion probabilities $P_{grn}$ and $P_{ign}$, and the time and scale conversion factors), the IGC kinetics can be simulated over a long time, for both stationary extreme conditions $C_{min}$ and $C_{max}$ (that is min($P_{grn}$, $P_{ign}$) and max($P_{grn}$, $P_{ign}$) respectively, see Figure 16(b)). Finally, the time dependence of the concentration $C(t)$ (Figure 16(a)) can be transposed into time dependent $P_{grn}(t)$ and $P_{ign}(t)$ (for example by linear interpolation between the min and max values). By implementing these time-dependent $P_{grn}(t)$ and $P_{ign}(t)$ in the CA model, the evolution of IGC can be simulated in the process (Figure 16(c)).

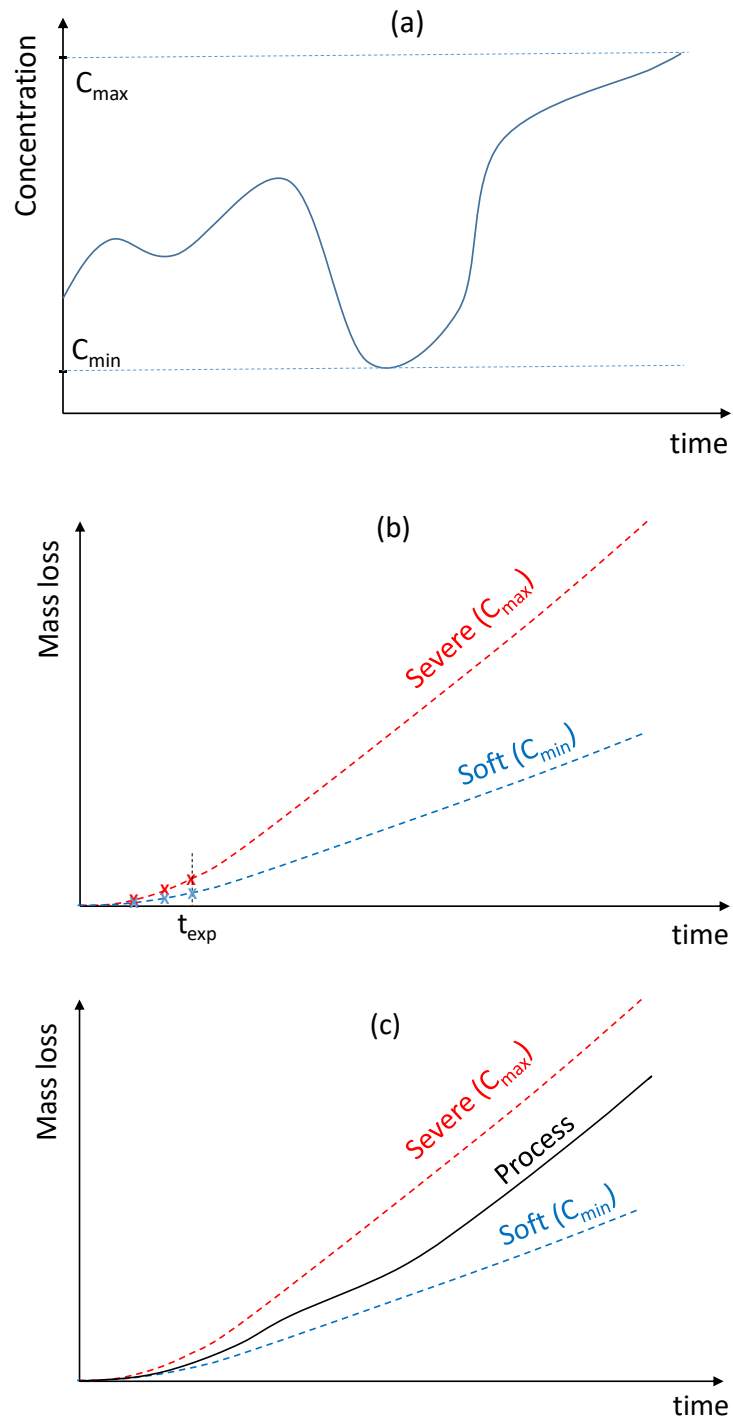

**Figure 16: Qualitative representation of the application of the CA model for simulating the long-term evolution of IGC in a case of a non-stationary process. (a) Expected medium aggressiveness in process. (b) Short time experiments (crosses) / long time CA simulations (dotted lines). (c) CA simulated mass loss in process (continuous line), that is bounded by the soft and severe mass losses.**

To go further in a more realistic description of IGC, we have seen that localized intragranular corrosion can also occur (Figure 4, Figure 5 or Figure 7). This kind of corrosion is probably due to the presence of linear (dislocations) or planar (twin boundaries…) defects that undergo a preferential attack. This kind of behavior could be implemented in the CA model by adding the corresponding states in the CA states list, generating these defects in the Voronoï diagram and finally by associating to them a specific corrosion probability.

**Conclusions**

By both an experimental and a CA modeling approaches, we investigated the IGC behavior of a SS in oxidizing medium. We showed that the CA model is able to simulate accurately the IGC behavior that is observed experimentally in all studied conditions. Therefore, we used the results of the CA simulations to quantify precisely the effect of the oxidizing medium on the IGC of the SS.

We first investigated the impact of the oxidizing character of the nitric medium on the evolution of the IGC of an austenitic SS. In two different stationary chemical conditions ("severe" and "soft", respectively), we showed that the SS reaches a different steady state in terms of the oxide and the geometrical natures of the solid interface with the medium: the oxide is thicker, the intergranular grooves are thinner and the surface area is larger with the oxidizing character of the nitric medium. Then we investigated the effect of an alternative "severe then soft" corrosion sequence. We showed that the system re-adapted to the "soft" conditions after a certain duration (which corresponds to the time necessary to corrode a thickness equivalent to the mean grain size of the SS) without memory effect from the previous "severe" conditions: the corrosion rate, the geometrical nature of the solid interface with the medium and the thickness of the oxide become similar to the system corroded in "soft" conditions only. In other words, whatever the IGC history of a SS, the new conditions of IGC corrosion need to consume the equivalent of a layer of grains to erase the effects of this history.

**Acknowledgments**

The authors thank Orano for financial support. They thank also R. Golchha for his careful proofreading. D. di Caprio would like to thank Dr. B. Diawara for fruitful discussions and introducing him to the marching cubes algorithm.

# Appendix

Appendix 1: solid/solution frontier estimation, marching square type algorithm

In this section, we detail the algorithm used to measure the length of the border line between the solution cells and the solid corresponding to GRN or IGN cells. For 3D objects, the marching cubes algorithm [62] is often used to determine contour lines. Given its simplicity, the algorithm finds applications in video games but also in medicine, for 3D visualization of volume data from magnetic resonance imaging and computed tomography scans [62, 63]. In the case of 2D images, the algorithm is referred to as marching squares. In the context of this paper, the algorithm is further adapted for a 2D grid based on rectangles paving the plane instead of squares. The details about the system coordinates used for the hexagonal lattice are given hereafter.

*The hexagonal lattice*

The plane of the hexagonal lattice structure is the horizontal 0xy plane, with, for $z = 0$, the first line $y = 0$, points starting at $x = 0:5$ and placed in the $Ox$ direction at a distance of 1 as shown in Figure 17 in blue points.

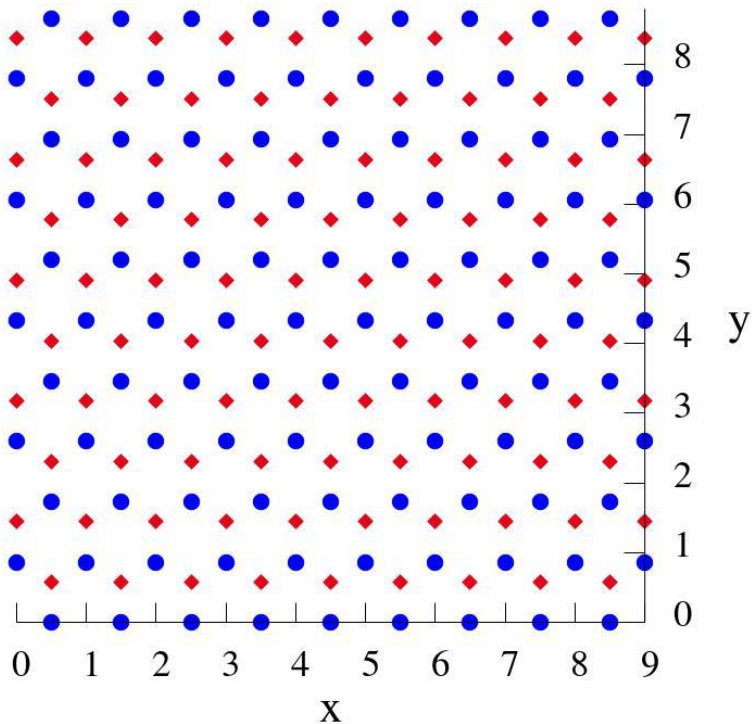

**Figure 17: Horizontal sections of the hexagonal lattice. Two consecutive planes, $z = 0$ and $z = (2/3)^{1/2}$, are shown with respectively blue and red points.**

Once these first two planes are defined, the rest of the lattice is derived. In particular, in the experiments and in the simulations vertical sections are studied. The layout of points in the vertical sections for the simulation is shown in Figure 18 where we have considered vertical *Oxz* planes with constant y coordinate.

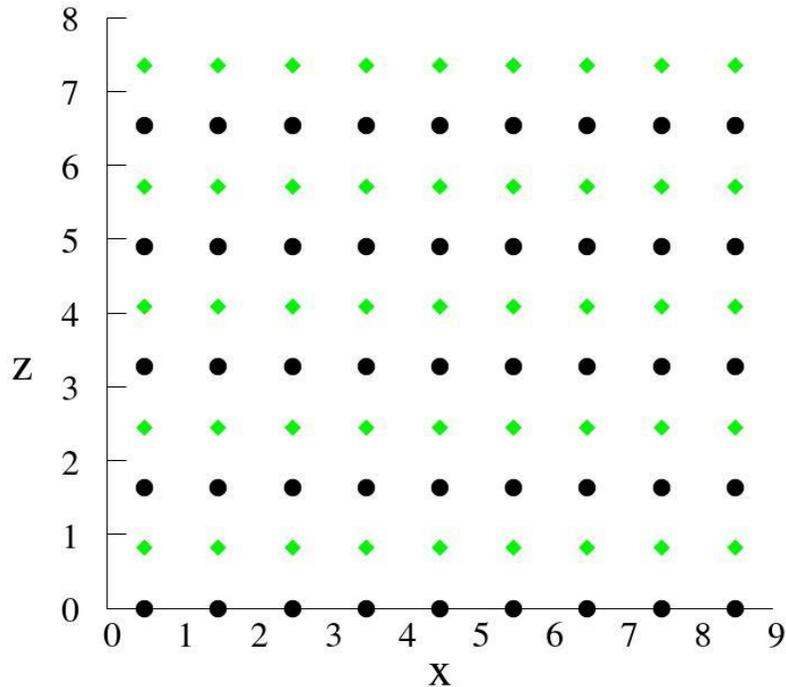

**Figure 18: Vertical sections of the lattice. Two consecutive planes for *y* = 0 and *y* = (3)$^{1/2}$/2 are shown with respectively black and green points.**

From the figure, it is clear that we have rectangles paving the plane.

*The marching rectangles algorithm*

Then, the algorithm consists in summing the contributions to the solution/solid border coming from each rectangle in the plane. For this aim, we first list all possible contributions coming from the configurations as shown in Figure 19. By assigning "1" to an intergranular or granular point and "0" to a solution point and starting from the bottom left corner, in an anticlockwise order, we obtain a binary number which is assigned to the corresponding case.

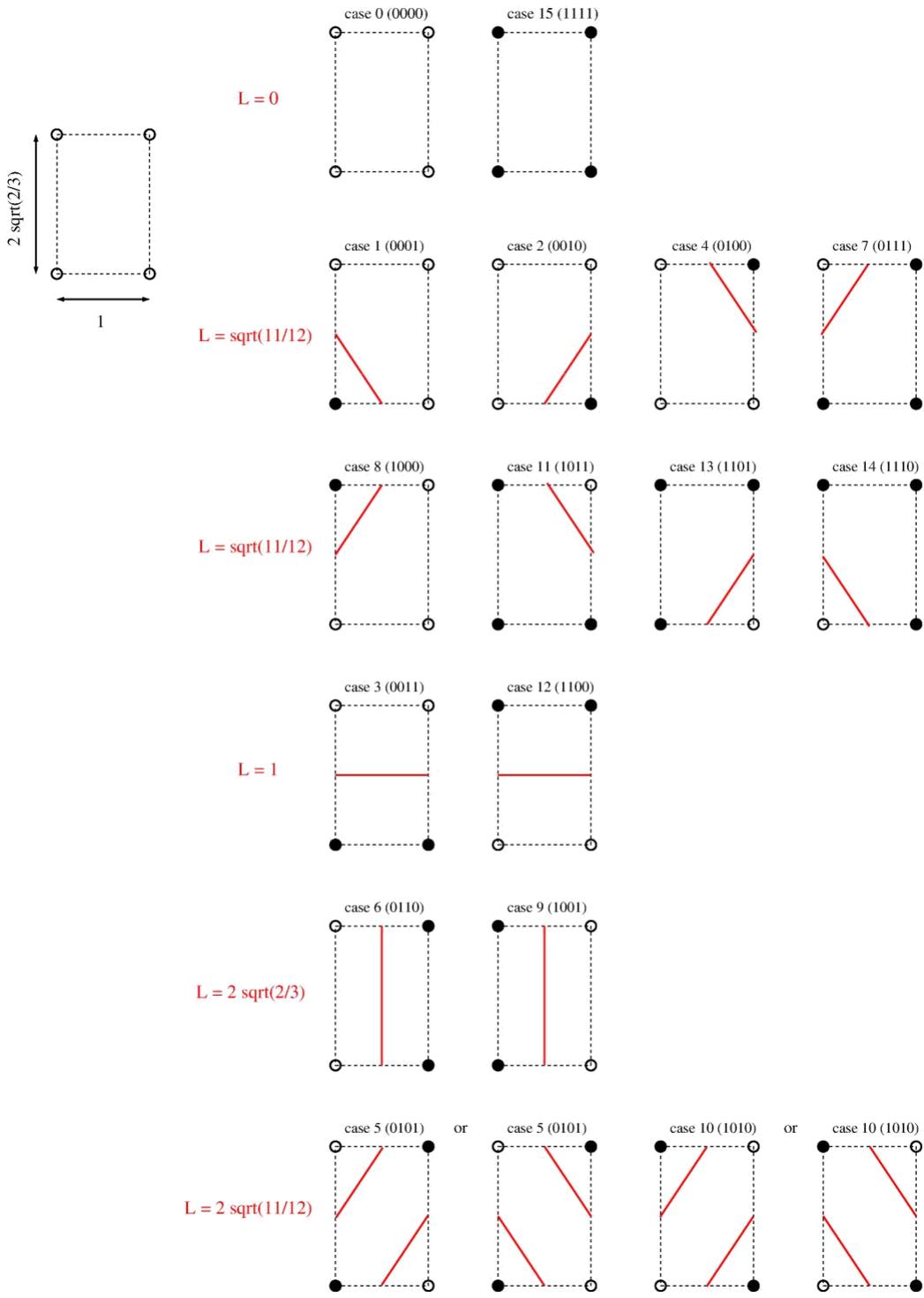

**Figure 19: On the top left the elementary rectangle and its dimensions. On the right are listed the length contributions and the corresponding layout of the points. Empty circles are for solution points and black points are for solid granular or intergranular cells.**

The cases in the first line correspond to cases inside the solution or inside the solid and they naturally do not contribute to the border length: $L = 0$. The cases in the last line correspond to a saddle point situation where we have two lines but for which the line orientation can be ambiguous. However, the contribution to the border length is the same for both orientations so that the result is unchanged despite the ambiguity. Other border lengths correspond to their geometrical contribution. All rectangles constituting the lattice are computed and their contribution is summed up which gives the final estimation of the border length. The ratio of this length divided by the straight lateral length of the system is what we consider to be the relative surface, which is in fact a line for the 2D sections.

Appendix 2: extracting the grain scale relative surface

Simulated sections of the system in Figure 5, Figure 7 and Figure 13 clearly show that we have two types of relative surface contributions. One with short length scale variations corresponding to the dissolution of a single grain and one at a larger scale, the scale of grains, due to IGC and detachment of grains. It is this last contribution that we are interested in. However, the algorithm presented above does not distinguish the two, as it provides the total separation line between solid and solution.

We present, hereafter, the method used to separate the contribution from IGC and grain detachment from the contribution at a short length scale due to dissolution. For this aim, we perform supplementary simulations for the dissolution of an exclusively granular material with an initial plane surface and in the identical conditions of "severe", "soft" and variable IGC conditions as for the IGC material model. Results are obtained averaging over 2000 sections. Figure 20 shows the relative surface, roughness for the three cases.

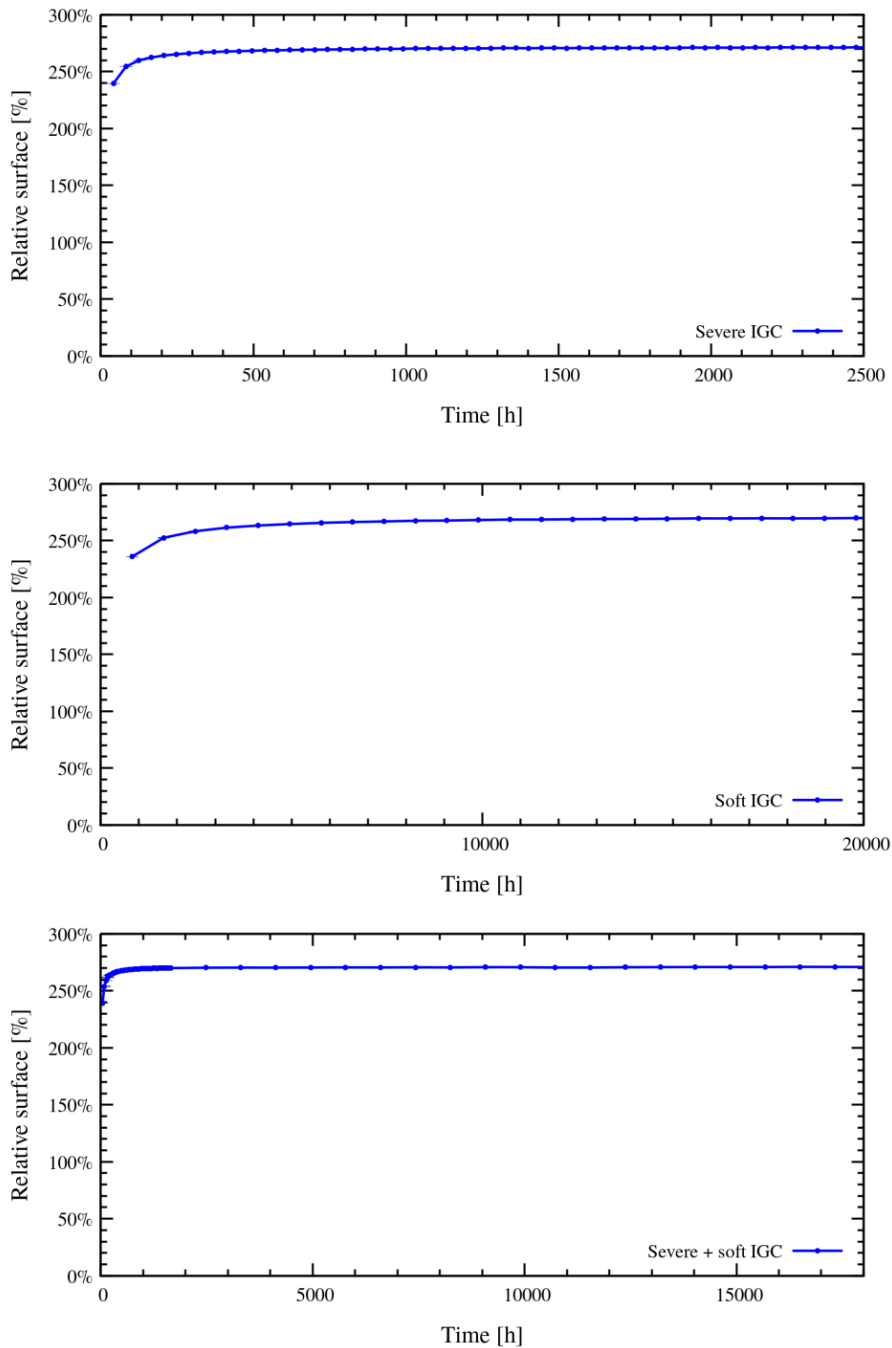

**Figure 20: Solid/solution relative interface for successively the "severe", "soft" and variable IGC conditions for granular dissolution.**

The relative surface varies with time: it increases before reaching a steady state. This is why we use this time dependent quantity to renormalize the relative surface and not the steady state value. Thus the relative surface corresponding to IGC and grain detachment is obtained from the relative surface obtained by the algorithm by dividing it by the relative surface for exclusively granular dissolution. The procedure has been tested by applying it to the study presented in [33], and the results compare well within the error bars. The results in the case of the present paper study are presented in Figure 10. Averages are performed using 100 Voronoï structures with for each 9 sections, this ensures that sections are sufficiently spaced to avoid undesirable correlations.